Journal of Big Data

# Big data analysis and distributed deep learning for next-generation intrusion detection system optimization

Khloud Al Jallad[*] , Mohamad Aljnidi and Mohammad Said Desouki

*Correspondence:
khloud.aljallad@hiast.edu.sy
Faculty of Information
Technology, Higher Institute
for Applied Sciences
and Technology, Damascus,
Syria

**Abstract**

With the growing use of information technology in all life domains, hacking has become more negatively effective than ever before. Also with developing technologies, attacks numbers are growing exponentially every few months and become more sophisticated so that traditional IDS becomes inefficient detecting them. This paper proposes a solution to detect not only new threats with higher detection rate and lower false positive than already used IDS, but also it could detect collective and contextual security attacks. We achieve those results by using Networking Chatbot, a deep recurrent neural network: Long Short Term Memory (LSTM) on top of Apache Spark Framework that has an input of flow traffic and traffic aggregation and the output is a language of two words, normal or abnormal. We propose merging the concepts of language processing, contextual analysis, distributed deep learning, big data, anomaly detection of flow analysis. We propose a model that describes the network abstract normal behavior from a sequence of millions of packets within their context and analyzes them in near real-time to detect point, collective and contextual anomalies. Experiments are done on MAWI dataset, and it shows better detection rate not only than signature IDS, but also better than traditional anomaly IDS. The experiment shows lower false positive, higher detection rate and better point anomalies detection. As for prove of contextual and collective anomalies detection, we discuss our claim and the reason behind our hypothesis. But the experiment is done on random small subsets of the dataset because of hardware limitations, so we share experiment and our future vision thoughts as we wish that full prove will be done in future by other interested researchers who have better hardware infrastructure than ours.

## Introduction

Recently, we have seen lots of real-life examples of attacks' huge impacts in different domains such as politics and economics. Hacking has become more critical and more dangerous than ever before. The number of hacking attacks is growing exponentially every few months. That means signature-based IDS is not useful anymore as we cannot update it with new signatures every few minutes. Also with developing technologies attacks become more sophisticated, APT attacks are more common than ever before.

Traditional IDS becomes inefficient. Other reasons why traditional IDS cannot support long-term, large-scale analytics as [1] said

1. Retaining large quantities of data wasn't economically feasible before. As a result, in traditional infrastructures, most event logs and other recorded computer activities were deleted after a fixed retention period (for instance, 60 days).
2. Performing analytics and complex queries on large, unstructured datasets with incomplete and noisy features, was inefficient. For example, several popular security information and event management (SIEM) tools weren't designed to analyze and manage unstructured data and were rigidly bound to predefined schemas. However, new big data applications are starting to become part of security management software because they can help clean, prepare, and query data in heterogeneous, incomplete, and noisy formats efficiently.
3. The management of large data warehouses has traditionally been expensive, and their deployment usually requires strong business cases. The Hadoop framework and other big data tools are now commoditizing the deployment of large-scale, reliable clusters and therefore are enabling new opportunities to process and analyze data.

## IDS types

IDS in general has three basic types based on its location: host IDS, network IDS and hybrid IDS, as shown in Fig. 1.

Network IDS is the domain of this experiment, so we will talk about in more details.

After deep research, we conclude NIDS Hierarchy shown in Fig. 2. NIDS has two basic types based on the data source that it is monitoring.

- *Log-based NIDS* that analyzes logs written by security devices when packets flow.
- *Raw data-based NIDS* that analyzes the data sent itself, it has two types

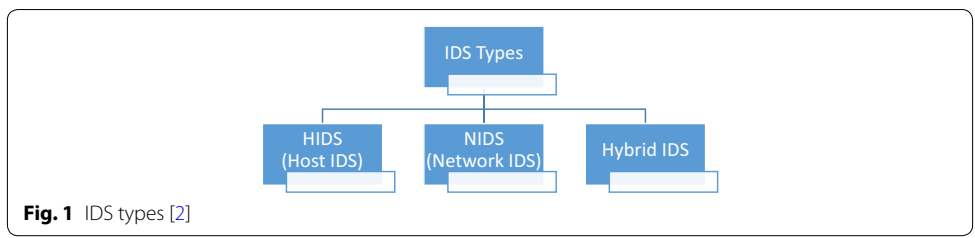

**Fig. 1** IDS types [2]

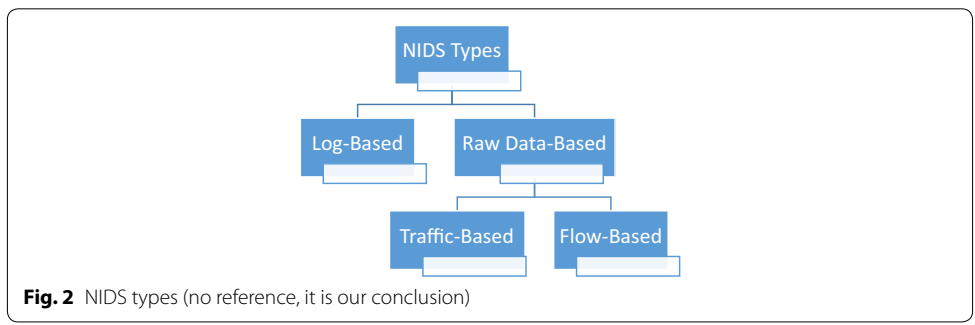

**Fig. 2** NIDS types (no reference, it is our conclusion)

- *Traffic-based* that contains the whole packets' data, headers and bodies.
- *Flow-based* that contains only headers of packets.

As for Traffic-based (packet-level NIDS), also called Deep Packet Inspection (DPI) or traditional packet-level NIDS, it is considered time-consuming when it comes to big data networks (more than 1 TB in second) or it will need a high cost of needed servers for just a very small optimization in performance, so we have to decide a tradeoff cost and accuracy. Some researchers choose to filter some packets to reduce costs [3].

As for Flow-based (flow-level NIDS), also considered Behavioral Analyzer NIDS, the body of each packet is ignored, only headers of packets are used to extract tuples. Each tuple has five values Source IP, Destination IP, Source Port, Destination Port, Protocol. Flow-level is better than packet-level in big networks when it comes to the cost of processing and storage, as it has very less cost because it processes only headers without bodies. Flow-based data approach is very lightweight. Storage issues that appeared in the packet-based approach are almost disappeared, but some types of attacks that are injected in bodies cannot be detected by analyzing headers only [3].

In general, Flow-level NIDS uses anomaly-based detection methods and packet-level NIDS uses signature-based detection methods. Each type has its advantages and disadvantages. Therefore, we need to tradeoff high cost or high false positive. Some researchers believe that a combination of both is the best solution [3].

The specific domain of this research paper is Flow-based. The reason behind choosing flow-based is that our used-dataset MAWI archives have headers only. Also, merging the flow-level of network behavior with anomaly-based gives the ability to detect new threats with lower cost, smaller storage and we decide to reduce false positive by using big data analysis.

## Anomaly types

Almost Anomaly-IDS detects only point anomalies. But anomalies have many types that need to be detected. Even it may be much more dangerous and more common.

Three basic categories of anomalies are [4].

I. *Point anomalies* (*outliers*), as shown in Fig. 3, often represent some extremum, irregularity or deviation that happens randomly and have no particular meaning. For example, a user trying to be root (U2R attack).

II. *Collective anomalies*, as shown in Fig. 4, often represented as a group of correlated, interconnected or sequential instances. While each particular instance of this group doesn't have to be anomalous itself, their collective occurrence is anomalous. For instance, Deny of Service attacks (DOS) are kind of collective anomalies as each request is normal by its own, but all together are considered anomaly.

III. *Contextual anomalies*, as shown in Fig. 5, represent an instance that could be considered as anomalous in some specific context. This means that observing the same point through different contexts will not always give us an indication of anomalous behavior. The contextual anomaly is determined by combining contextual and behavioral features. For contextual features, time and space are most frequently used. While the behavioral features depend on the domain that is being analyzed, ex. amount of money spent, average temperature, number of bytes, number of packets, or any other quantitative measure. Security domain has lots of examples of contextual anomalies that almost available IDS cannot detect.

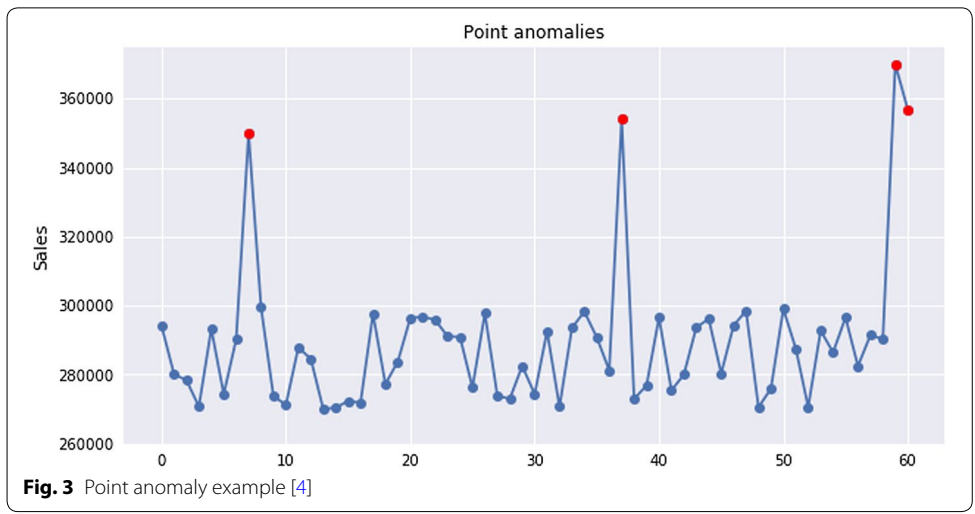

**Fig. 3**  Point anomaly example [4]

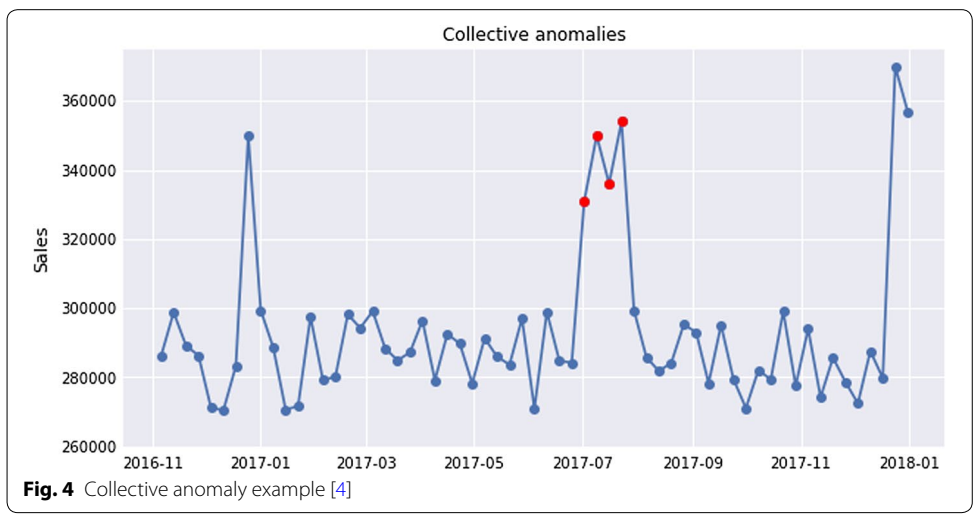

**Fig. 4**  Collective anomaly example [4]

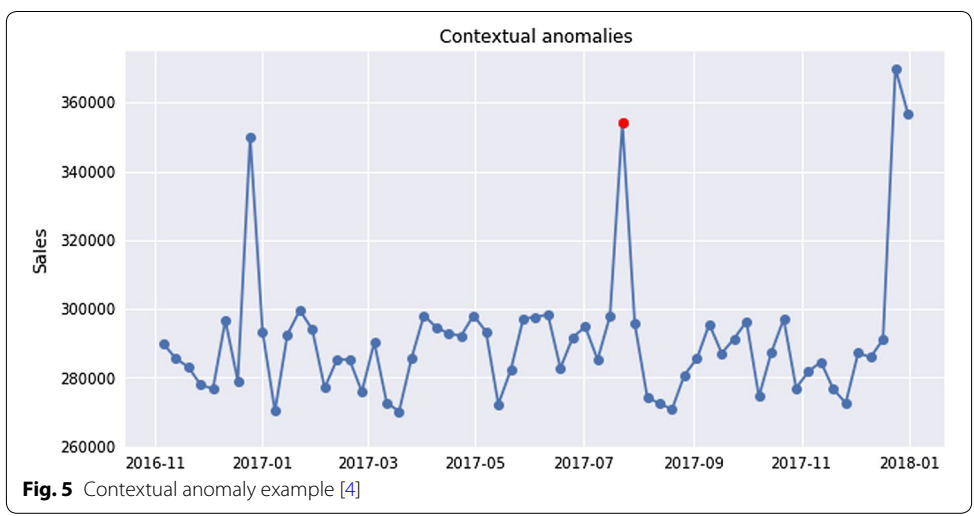

**Fig. 5**  Contextual anomaly example [4]

**Table 1  Examples of Network Anomalies Types [5]**

| Anomaly | Network Attack |
| --- | --- |
| Point | U2R, R2L |
| Contextual | Scanning (Probe) |
| Collective | DoS |

Table 1 contains examples of network anomalies types [5].

## Related work

We have two basic types of researches in this domain, researches of traditional IDS optimization and researches of using big data for IDS optimization.

### Traditional IDS optimization

Table 2 contains research experiments that used basically SVM and optimize it by adding another model to its result or to its input [2].

### IDS optimization using big data

We have two basic categories of these researches. First, researches that just proposed the idea of using big data for optimization. It expected promising results without doing actual experiments or having any proof of the idea as it was just a suggestion of a general model. The second type of researches contains real experiments that was done trying to prove these suggests. Some of them apply SVM as it was the best in the traditional domain, others propose that may another algorithm will be better than SVM in big data environment. Some of experiments focus on used big data tools, such as spark or storm, without applying any mining algorithm, just a simple threshold.

#### General model

Many papers proposed using big data on security for optimization and unknown attacks detection [17, 18], as shown in Fig. 6. Since then next generation anomaly security systems using big data has been a hot research topic domain as it is promising to be one of the optimal solutions for hacking detection problem.

#### Done experiments

Table 3 contains a comparison of some done research papers in this domain as per [19].

## Data

The dataset used for this experiment is a combination of three datasets: Flows extracted from MAWI Archives, labels from MAWILAB and aggregated flows from AGURIM as shown in Fig. 7.

**Table 2  Comparisons between researches that used SVM as basic classifier**

| Approach | Working idea | Dataset | Detection rate | False alarm |
|---|---|---|---|---|
| Unsupervised anomaly detection system [6] | Tune and optimize automatically the values of parameters without pre-defining them | From Kyoto University honeypot | – | – |
| Multiclass SVM [7] | Attributes are optimized using k-fold cross validation. This technique can be used to decrease the rate of False-Negatives in the IDS | Self | – | – |
| OC-SVM One-Class SVM [8] | Multistage OC-SVM and feature extraction represents a method to detect unknown attacks. Method is poor in second stage classifier to detection rate of unknown attacks | From Kyoto | 80.00 | 20.94 |
| IG-ABC-SVM Information Gain-Artificial Bee Colony [9] | A combining IG feature selection and SVM classifier in IDS model is proposed. Experiments using just two swarm intelligence algorithms | NSL-KDD | 98.53 | 0.03 |
| SbSVM [10] | Autonomous labeling algorithm of normal traffic (when the class distribution is not imbalanced). Not evaluated for real-time case | DARPA | 99 | 5.5 |
| RS-ISVM-reserved set-incremental SVM [11] | An incremental SVM training algorithms is used, hybrid with modifying kernel function U-RBF Foreseeing attacks, specifically for attacks of U2R and R2L may not tolerate but oscillation problem solved | KDD Cup 1999 | 89.17 | 4.9 |
| SVM-GA [12] | Hybrid model by combining (GA and SVM) | KDD CUP 1999 | 98.33 | 0.50 |
| Genetic principal component [13] | Subset selection using GA and PCA | KDD cup 1999 | 99.96 | 0.49 |
| SVM and NN [14] | Hybrid process. Most significant performance as far as training time but time consuming and hard task to trigger | DARPA | 99.87 | – |
| N-KPCA-GA-SVM kernel PCA genetic algorithm-SVM [15] | Hybrid of KPCA, SVM and GA algorithms. Faster convergence speed. Performs higher predictive accuracy and better generalization. But have complex structure and have latency for real-time application | KDD CUP99 | 96.37 | 0.95 |
| CSV-ISVM Candidate Support Vector-Incremental SVM [16] | Improved learning algorithm to better recognize rate and false alarm rate than usual classification | KDD Cup 1999 | 90.14 | 2.31 |

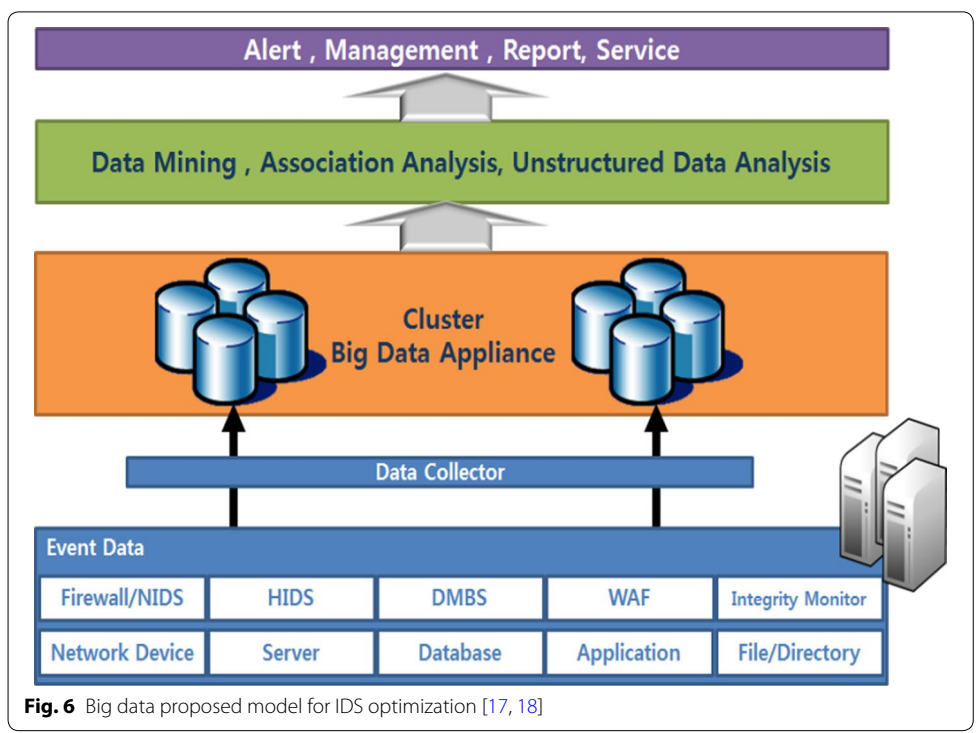

**Fig. 6** Big data proposed model for IDS optimization [17, 18]

As we are working on contextual attacks detection, we form a network traffic language that is a series of traffic and traffic aggregation information, that includes each second statistics with the second before and the second after. Thus, the resulted system will have a delay of one second. Each second information is a combination of the three data sources: MAWI, MAWILAB, and AGURIM.

**MAWI**

MAWI [41] stands for Measurement and Analysis on the WIDE Internet [42–44]. It is a new real-life dataset that is publically available for free. It contains real-life traffic data of Japan-US cable. It is collected and preprocessed by a sponsor of the Japanese ministry of communication. For privacy concerns, all critical information is replaced by other values and all packet loads are removed.

We use MAWI traffic to extract flow information with statistics about flow. For example, Source IP, Destination IP, Frame Length, IP Length, IP Version, TTL Window Size, Flags … etc.

**MAWILAB**

MAWILAB [45] is a project done on top of MAWI archive that contains labels of data, and it is updated automatically every day. Labeling data is done by community of four classifiers. Classifiers are Principal component analysis (PCA), Gamma Distribution, Hough Transform, Kullback–Leibler (KL). Labels are tagged according to class of majority classifiers detection. That help reducing false positive rate.

Labels are done according to taxonomy of anomalies in network traffic as shown in Fig. 8 [46].

**Table 3  Done experiments on IDS optimization using big data**

| Name | OSI layer | Time of detection | Data sources | Detection technique | Big data environment | Used data set |
|---|---|---|---|---|---|---|
| Beehive [20] | 7 | Non-real | Proxy logs | k-means | Hadoop, Hive | Operational network |
| Bumgardner and Marek [21] | 3, 4 | Real-time | Network flows | Threshold | Storm, HBase, Hadoop | Operational network |
| Camacho et al. [22] | 7, 4, 3 | Non-real | Firewall and IDS logs | PCA | Custom | Public dataset |
| Dromard et al. [23] | 4, 3 | Non-real | Network flows | DBSCAN | Spark | Operational network |
| Giura and Wang [24] | 7, 4, 3 | Non-real | Network and application data | Threshold | Hadoop | Operational network |
| Gupta and Kulariya [25] | 7, 4, 3 | Non-real | Network captures | Several feature extraction and clas-sification algorithms | Spark | Public dataset |
| Gonçalves et al. [26] | 3, 4, 7 | Non-real | DHCP, Authentica-tion and Firewall logs | EM | Hadoop, Weka | Operational network |
| Hashdoop [27] | Packet cap-tures | Non-real | Network traffic | | Hadoop | Public dataset |
| Iturbe et al. [19] | | Non-real | Network flows | Whitelisting | Elastics Search | Operational network |
| Marchal et al. [28] | 3, 4, 7 | Non-real | Honeypot, DNS, HTTP, Network flow data | Threshold | Hadoop, Hive, Pig, Spark | Operational network |
| MATATABI [29] | 3, 4, 7 | Non-real | DNS records, Network flows, Spam email | Multiple clas-sification algorithms | Hive | Operational network |
| Rathore et al. [30] | 3, 4 | Non-real | Network flows | C4.5, RepTree | Spark, Weka | Public dataset |
| Ratner and Kelly [31] | Packet cap-tures | Non-real | Network packets | Manual data querying | Hadoop | Operational network |
| Therdphapi-yanak and Piromsopa [32] | 7, 4, 3 | Non-real | Network logs | k-means | Hadoop, Mahout | Public dataset |
| TADOOP [33] | 3, 4 | Non-real | Network flows | DTE-FP | Hadoop | Operational network |
| Wang et al. [34] | 3, 4 | Real-time | Network flows | Intergroup entropy, LMS | Storm | Operational network |
| Xu et al. [35] | 7 | Non-real | Console logs | PCA | Hadoop | Operational network |
| Hadzios-manovic et al. [36] | | Non-real | SCADA logs | FP-Graph | Custom | Operational network |
| Difallah et al. [37] | | Real-time | Process data | LISA | Storm | Simulated process data |
| Wallace et al. [38] | | Real-time | Process data | Cumulative probability distribution | Spark | Operational network |

**Table 3 (continued)**

| Name | OSI layer | Time of detection | Data sources | Detection technique | Big data environment | Used data set |
|------|-----------|-------------------|--------------|---------------------|----------------------|---------------|
| Kiss et al. [39] | | Non-real | Process data | *k*-means | Hadoop | Simulated process data |
| Hurst et al. [40] | | Non-real | Process data | Multiple classification algorithms | Custom | Simulated process data |

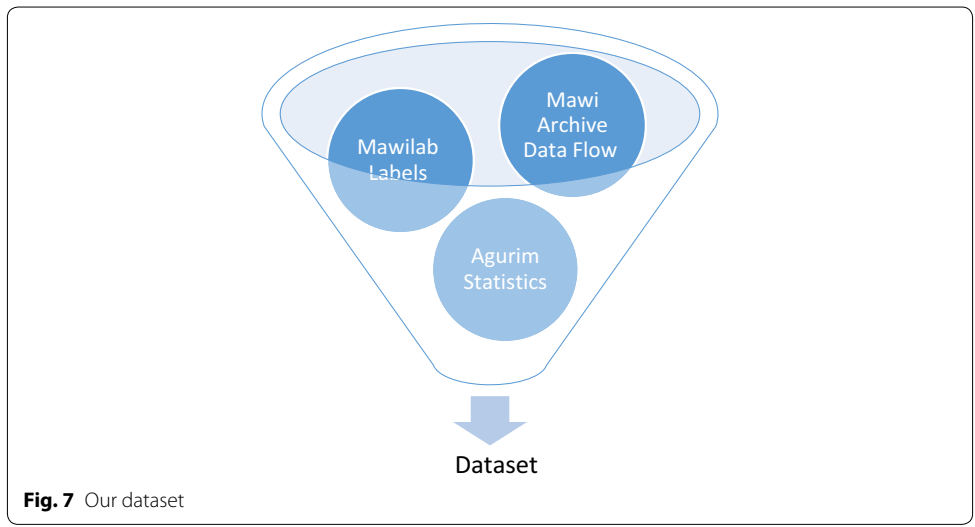

**Fig. 7** Our dataset

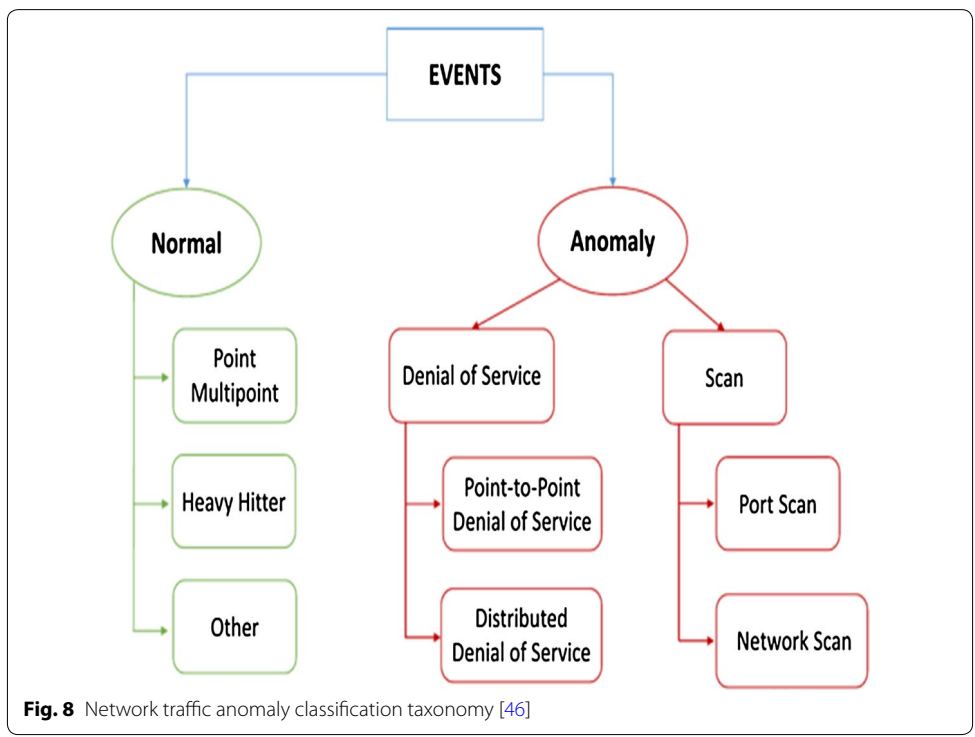

**Fig. 8** Network traffic anomaly classification taxonomy [46]

### MAWILAB meta data

As per [47], Anomalies are reported in CSV format. Each line in the CSV files consists of a 4-tuple describing the traffic characteristics and additional information such as the heuristic and taxonomy classification results. The actual order of the fields is given by the CSV files header:

*anomalyID, srcIP, srcPort, dstIP, dstPort, taxonomy, heuristic, distance, nbDetectors, label*

anomaly ID is a unique anomaly identifier. Several lines in the CSV file can describe different sets of packets that belong to the same anomaly. The anomaly ID field permits to identify lines that refer to the same anomaly.

All fields are

- srcIP is the source IP address of the identified anomalous traffic (optional).
- srcPort is the source port of the identified anomalous traffic (optional).
- dstIP is the destination IP address of the identified anomalous traffic (optional).
- dstPort is the destination port of the identified anomalous traffic (optional).
- Taxonomy is the category assigned to the anomaly using the taxonomy for backbone traffic anomalies.
- Heuristic is the code assigned to the anomaly using simple heuristic based on port number, TCP flags and ICMP code.
- Distance is the difference Dn−Da
- nbDetectors is the number of configurations (detector and parameter tuning) that reported the anomaly.
- Label is the MAWILab label assigned to the anomaly, it can be either: anomalous, suspicious, benign, or notice.

Labels are

- Abnormal, if all four classifiers consider it as an attack.
- Benign, if all four classifiers consider it as normal.
- Suspicious, if three out of four classifiers consider it as an attack.
- Notice, if three out of four classifiers consider it normal.

We use MAWILAB for labeling traffic data. The output of our Chatbot is a language that has two words only, Anomaly and Benign. Labels for output are extracted from MAWILAB four labels where we consider the majority classifier result.

- Abnormal, suspicious

We consider it Anomaly because the majority of classifiers detect it as an attack.

- Benign, notice

We consider it as Benign because the majority of classifiers detect it as normal.
The reason behind choosing the majority results is overcoming false positive problem.

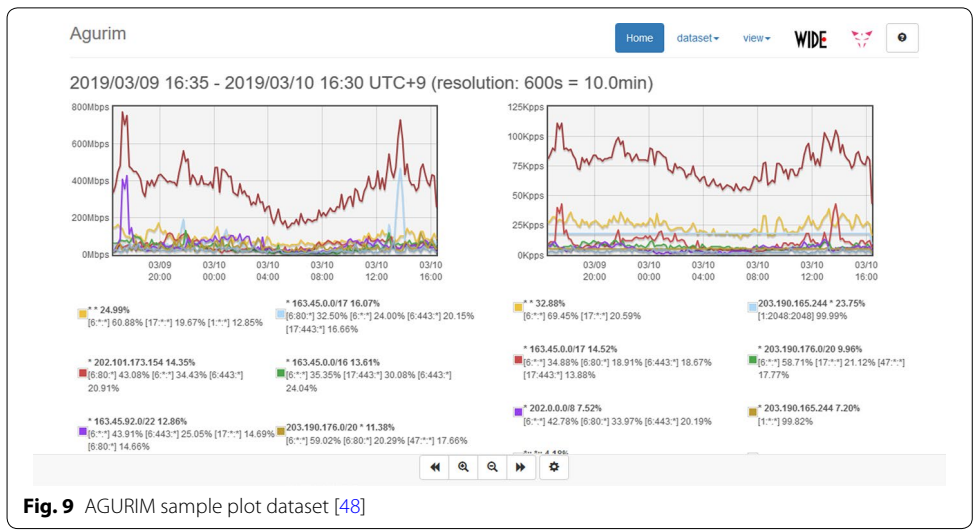

**Fig. 9** AGURIM sample plot dataset [48]

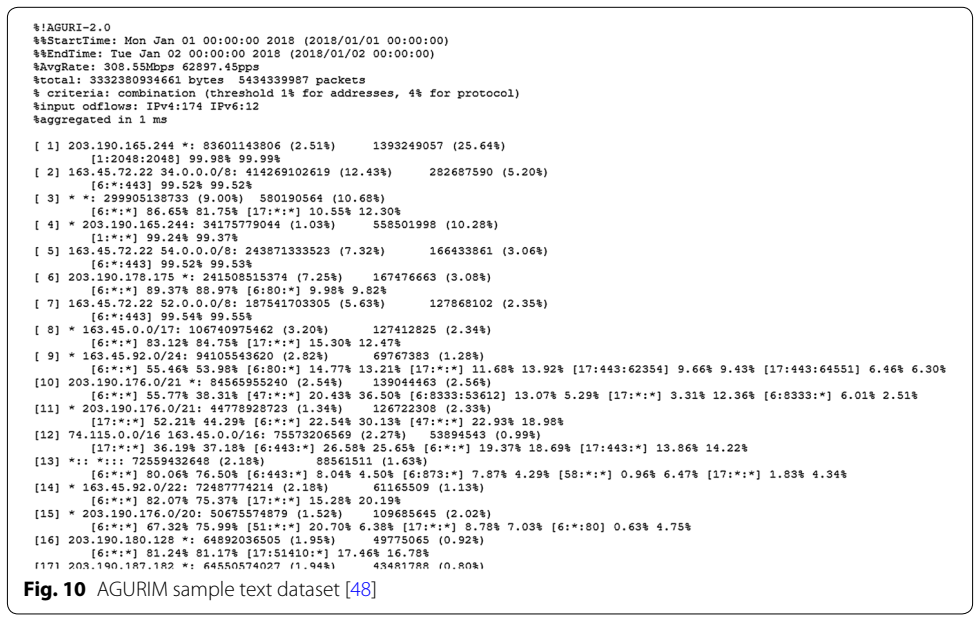

**Fig. 10** AGURIM sample text dataset [48]

## AGURIM

AGURIM is a project done on MAWI archive traces which is a network traffic monitor based on flexible multi-dimensional flow aggregation in order to identify significant aggregate flows in traffic. It has two views, one based on traffic volume and the other based on packet counts, address or protocol attributes, with different temporal and spacial granularities. The supported data sources are pcap, sFlow, and netFlow [48–50]. Data has two formats, Texts and plots as you can see in Figs. 9, 10.

*AGURIM meta data*

Each instance of data is represented by two lines. The first line of an entry shows the information of the source–destination pair: the rank, source address, destination address, percentage in volume, and percentage in packet counts. The second line shows the protocol information within the source destination pair protocol, source port, destination port, percentage in volume, and percentage in packet counts. A wild-card, "*", is used to match any.

## Method

We propose Networking Chatbot, a deep recurrent neural network: using Long Short Term Memory (LSTM) on top of Apache Spark Framework that has an input of flow traffic and traffic aggregation. The output is a language of two words, normal or abnormal. We propose merging the concepts of language processing, contextual analysis, distributed deep learning, big data, anomaly detection of flow analysis. We want to detect point, collective and contextual anomaly by creating a model that describes the network abstract normal behavior, as shown in Fig. 11.

Using big data analysis with deep learning in anomaly detection shows excellent combination that may be optimal solution. Deep learning needs millions of samples in dataset and that is what big data handle and what we need to construct big model of normal behavior that reduces false positive rate to be better than small anomaly models.

Using big data with time series will allow us to analyze bigger periods than before and utilizing it in IDS domain may allow to detect advanced threats that remains undetected in system too long, for months or may years. Because APT attacks happens slowly, analyzing 90 days only of traditional IDS is not enough to detect those collective anomalies.

Also, using big data with context of time frames will allow to detect contextual anomalies that were not possible to detect by traditional IDS.

System consists of two parts, feature extraction and classification.

### Feature extraction

We get the PCAP files from MAWI archive then extract flow statistics and label them by merging them with MAWILAB gold standard labels. Also, final resulted dataset is merged with flow aggregation from AGURIM dataset so that we can create a sequence of time frames, a frame of one second is used in this experiment. The idea behind using data flow aggregation is to deal with time frames of flows like we deal with time frames in videos that we compare each scene with its neighbor frames by their gradients. Furthermore, we can add a level of abstraction of network behavior by adding data aggregation to neural network input. This data set is used for training and testing.

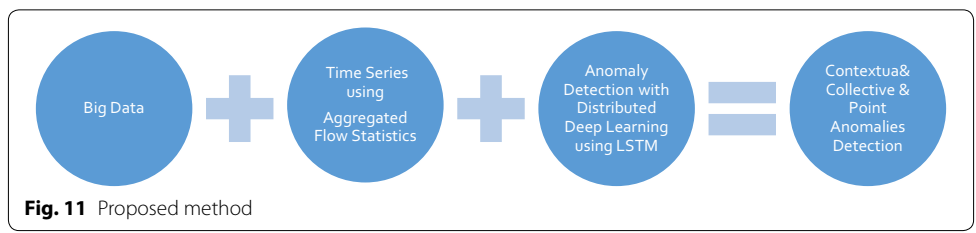

**Fig. 11** Proposed method

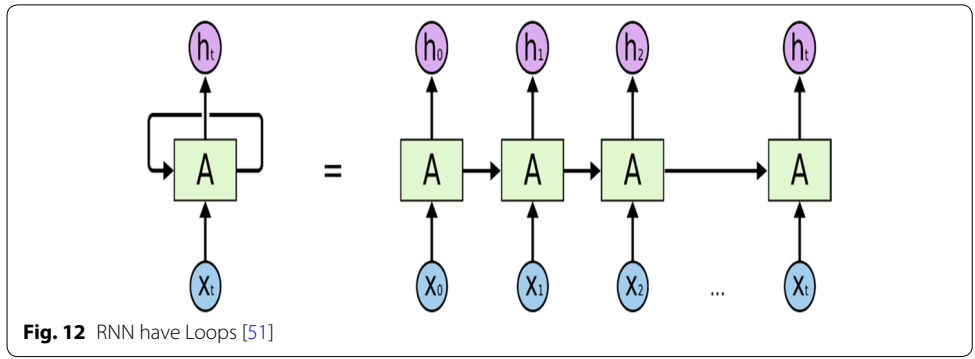

**Fig. 12** RNN have Loops [51]

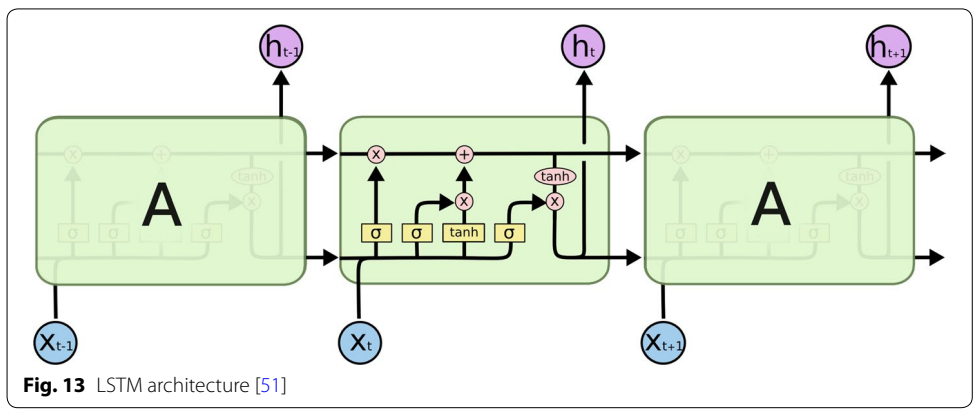

**Fig. 13** LSTM architecture [51]

### Classification

Choosing distributed deep neural network has been used by many different researchers as the assumption that it will optimize results by modeling millions of samples of data and more complicated neural networks with more options.

The reason behind choosing RNN is its ability to deal with sequences. RNN is extension of a convention feed-forward neural network. Unlike feedforward neural networks, RNN have cyclic connections making them powerful for modeling sequences. As a human no one think of each event separately. For example, when you are reading this article you read each word but you understand it in the context of this article, so that you understand the whole concept of this paper. That is the idea of RNN that has loops to deal with input as a sequence, and that what we need to handle each event on network within its context, as shown in Fig. 12.

Long Short Term Memory is a special case of RNN that solves problems faced by the RNN model [51, 52].

1. Long term dependency problem in RNNs.
2. Vanishing Gradient and Exploding Gradient.

Long Short Term Memory is designed to overcome vanishing gradient descent because it avoids long-term dependency problem. To remember information for long periods of time, each common hidden node is replaced by LSTM cell. Each LSTM cell consists of three main gates such as input gate *it*, forget gate *ft*, and output gate *ot*. Besides *ct* is cell state at time *t*.

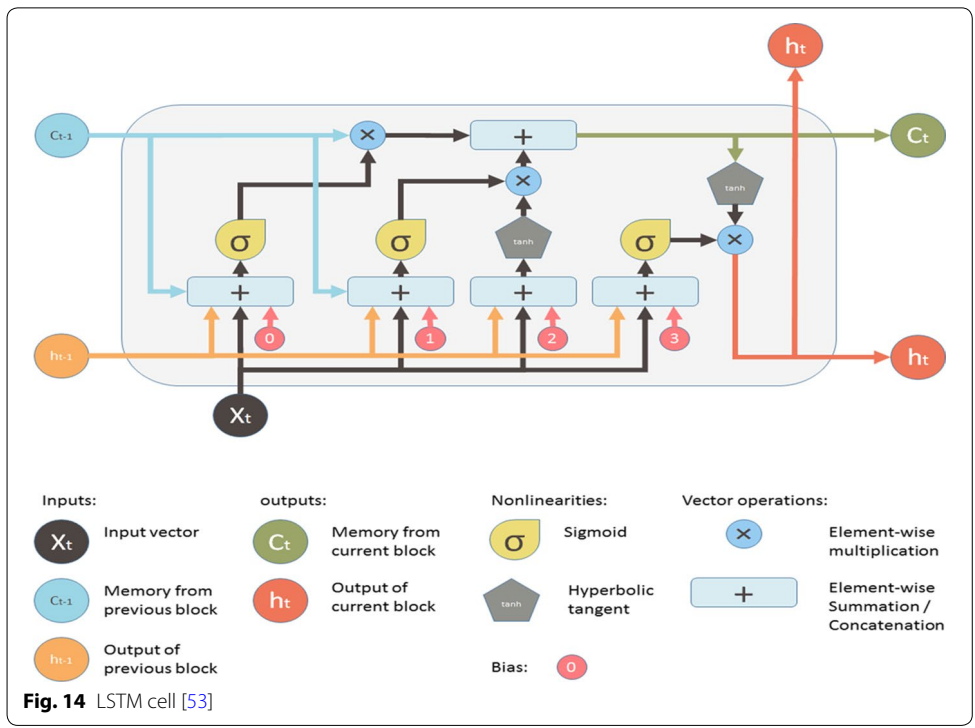

**Fig. 14** LSTM cell [53]

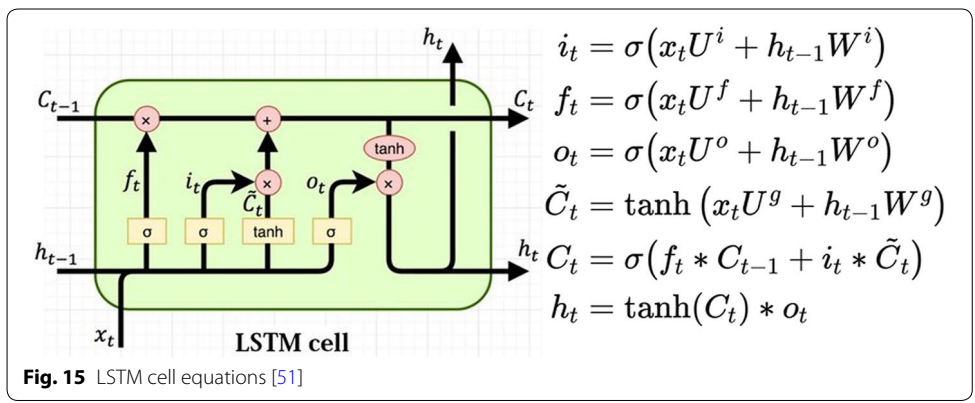

**Fig. 15** LSTM cell equations [51]

Long Short Term Memory architecture is shown in Figs. 13, 14.

The equations to calculate the values of gates is shown in Fig. 15, where *xt*, *ht*, and *ct* correspond to input layer, hidden layer, and cell state at time *t*. Furthermore, $\sigma$ is sigmoid function. Finally, *W* is denoted by weight matrix.

The reason behind choosing big data with anomaly detection is our interest in anomaly detection advantage of detecting new threats and our goal to reduce anomaly detection disadvantage of high false positive by training model with more normal samples.

The reason behind choosing deep learning with big data is the need for millions of samples with high number of feature sequences for training and that what big data systems can handle and that we propose to optimize normal model by training model with more normal cases more context features to reduce false positive without facing the problem of overfitting as fast as in traditional learning.

The reason behind choosing LSTM is our interest in contextual anomalies. So we propose a network language as an input for our Chatbot. Each sentence in the network language includes time series of packets flows, flows aggregations and statistics of each second with second before and second after.

## Proposed framework and libraries

### Colab

Colaboratory is a free research tool offered by Google, for machine learning education and research. It's a Jupyter notebook environment that requires no setup to use. Code is written on browser interface. Code is executed in a virtual machine dedicated to user account (options available now are CPU, Graphical Processing Unit GPU, Tensor Processing Unit TPU) [54].

### ELephas

Elephas brings deep learning with Keras [55] to Spark [56]. Elephas intends to keep the simplicity and high usability of Keras, thereby allowing for fast prototyping of distributed models (distributed deep learning [57]), which can be run on massive data sets [58].

### Experiment setup

This experiment is performed on Google CoLab [54] using Keras library (Python Deep Learning library) [55]. We apply LSTM of 64 hidden nodes with Relu activation function and dropout=0.5. Using binary cross-entropy loss function. Using RMSprop optimizer. Learning rate=0.001, rho=0.9, decay=0.0. Colab has space limitations in addition to execution time limitations, that was the reason that we prove only point anomalies but cannot prove collective and contextual anomalies.

## Results and discussion

We wanted to prove five aspects of results improvements, but we were able to prove only three of them because of hardware limitations. We were trying last year very seriously to execute the experiment on better hardware with no hope. We are in Syria and we have financial prohibition.

Although war and difficult conditions, we want to contribute to research. Therefore, we choose to share this paper with you, to share the insights we got and approximated results of average of experiments we did. And we wish that the complete experiment will be done in future by interested researchers.

Because of hardware limitations, experiments are done on random subsets of dataset. Therefore, we will talk about insights and an approximate percentage of all done experiments on average. We will not mention numbers and charts as it is not the exact one. We got different percentages for each experiment as it is random samples so numbers are not too accurate to provide them as charts or so.

We do not have enough hardware to test contextual anomalies and collective anomalies for long times. We experiment only point anomalies. Results we get by experiment shows that the accuracy of distributed contextual flow Chatbot model is higher than the accuracy of traditional learning model. False positive is getting lower by 10% less than traditional learning model. We used SVM to compare with, as it has one of the highest results among traditional learning classifiers.

Adding flow aggregation information to features, in addition to flow statistics information, is a good choice that increases accuracy and better describes the abstract behavior on a network. We can get use of gradients between each second and the seconds before and after. Also, adding context to be taken into consideration causes delay time equal to time frame taken in context.

Merging big data with anomaly detection with deep learning is an optimal solution that solves the problem of overfitting that causes high false positive. It allows us to detect new threats by anomaly methods with lower false positive by extending dataset of training to include more normal cases and much more features without facing the problem of overfitting as traditional learning.

## Conclusion and future work

Using big data with deep anomaly-IDS is promising in next-generation IDS because of its ability to detect new threats in different contexts with lower false positive than already used IDS.

We propose a model to analyze sequences of flows and flows aggregation for each second with seconds before and after. The experiment shows lower false positive, higher detection rate and better point anomalies detection. As for proof of contextual and collective anomalies detection, we discuss our claim the reason behind our hypothesis but we were not able to do complete experiment because of hardware limitations. The experiments we did on random small subsets of dataset were promising but not enough to prove our hypothesis. Therefore, we share experiment and our future vision thoughts of contextual and collective anomalies detection. We wish that complete experiment will be done in future by other interested researchers who have better hardware infrastructure than ours.

**Abbreviations**
IDS: Intrusion Detection System; RNN: recurrent neural networks; LSTM: Long Short Term Memory; FP: false positive; DoS: Deny of Service attack; R2L: remote to user; U2R: user to root.

**Acknowledgements**
I would like to thank my parents for their endless love and support.

**Authors' contributions**
KAJ took on the main role so she performed the literature review, conducted the experiments and wrote the manuscript. MJ and MSD took on a supervisory role and oversaw the completion of the work. All authors read and approved the final manuscript.

**Funding**
The authors declare that they have no funding.

**Availability of data and materials**
All datasets in this survey are available online, you can find links in references.

**Ethics approval and consent to participate**
The authors Ethics approval and consent to participate.

**Consent for publication**
The authors consent for publication.

## Publisher's Note